\documentclass[conference]{IEEEtran}
\IEEEoverridecommandlockouts
\usepackage{cite}
\usepackage{amsmath,amssymb,amsfonts}
\usepackage{algorithmic}
\usepackage{graphicx}
\usepackage{textcomp}
\usepackage{xcolor}
\usepackage{subcaption}
\usepackage{easyReview}
\usepackage{array}
\usepackage{float}

\def\BibTeX{{\rm B\kern-.05em{\sc i\kern-.025em b}\kern-.08em
    T\kern-.1667em\lower.7ex\hbox{E}\kern-.125emX}}

\usepackage[acronym,nopostdot,  style=super, nonumberlist, toc]{glossaries}
\newacronym[plural=HITLs,firstplural=human-in-the-loop (HITL)]{HITL}{HITL}{human-in-the-loop}
\newacronym[plural=CPSs,firstplural=cyber-physical systems (CPS)]{CPS}{CPS}{cyber-physical system}
\newacronym[]{TSN}{TSN}{time sensitive networking}
\newacronym[plural=CPSs]{UE}{UE}{user equipment}
\newacronym{HARQ}{HARQ}{Hybrid-ARQ}
\newacronym{MCS}{MCS}{modulation coding scheme}
\newacronym{MLP}{MLP}{multi-layer perceptron}
\newacronym{SDR}{SDR}{software-defined radio}
\newacronym{OAI}{OAI}{OpenAirInterface}
\newacronym{SINR}{SINR}{signal to interference and noise ratio}
\newacronym{CDF}{CDF}{cumulative density function}
\newacronym{CDE}{CDE}{conditional density estimation}
\newacronym{GMM}{GMM}{Gaussian mixture model}
\newacronym{GMEVM}{GMEVM}{Gaussian mixture extreme value model}
\newacronym{TDD}{TDD}{time division duplex}
\newacronym[plural=PRBs,firstplural=physical resource blocks (PRB)]{PRB}{PRB}{physical resource block}
\newacronym[plural=RANs,firstplural=radio access networks (RAN)]{RAN}{RAN}{radio access network}
\newacronym{COTS}{COTS}{commercial off-the-shelf}
\newacronym[plural=MDNs,firstplural=mixture density networks (MDN)]{MDN}{MDN}{mixture density network}
\newacronym[plural=MLEs,firstplural=maximum likelihood estimation (MLE)]{MLE}{MLE}{maximum likelihood estimation}
\newacronym[plural=KL-divergences,firstplural=Kullback-Leibler divergences (KL-divergence)]{KL-divergence}{KL-divergence}{Kullback-Leibler divergence}
\newacronym[]{EVT}{EVT}{extreme value theory}
\newacronym[plural=GPDs,firstplural=generalized Pareto distributions (GPD)]{GPD}{GPD}{generalized Pareto distribution}
\newacronym[plural=PDFs,firstplural=probability density functions (PDF)]{PDF}{PDF}{probability density function}

\usepackage{amsmath,amsfonts} 
\usepackage{graphicx}
    
\begin{document}





\title{Data-Driven Latency Probability Prediction for Wireless Networks: Focusing on Tail Probabilities\\
\thanks{This work was supported by the European Commission through the H2020 project DETERMINISTIC6G (Grant Agreement no. 101096504).}
}

\author{
    \IEEEauthorblockN{Samie Mostafavi, Gourav Prateek Sharma,
    James Gross}
    \IEEEauthorblockA{\text{KTH Royal Institute of Technology},Stockholm, Sweden \\
    \{ssmos,gpsharma,jamesgr\}@kth.se}
}

\maketitle

\begin{abstract}
With the emergence of new application areas, such as cyber-physical systems and human-in-the-loop applications, there is a need to guarantee a certain level of end-to-end network latency with extremely high reliability, e.g., 99.999\%.
While mechanisms specified under IEEE 802.1as time-sensitive networking (TSN) can be used to achieve these requirements for switched Ethernet networks, implementing TSN mechanisms in wireless networks is challenging due to their stochastic nature. 
To conform the wireless link to a reliability level of 99.999\%, the behavior of extremely rare outliers in the latency probability distribution, or the tail of the distribution, must be analyzed and controlled.
This work proposes predicting the tail of the latency distribution using state-of-the-art data-driven approaches, such as mixture density networks (MDN) and extreme value mixture models, to estimate the likelihood of rare latencies conditioned on the network parameters, which can be used to make more informed decisions in wireless transmission.
Actual latency measurements of IEEE 802.11g (WiFi), commercial private and a software-defined 5G network are used to benchmark the proposed approaches and evaluate their sensitivities concerning the tail probabilities.
\end{abstract}


\section{Introduction}

Traditionally, communication networks have been designed to provide best-effort connectivity between application endpoints without guaranteeing performance. 
In recent years, new application areas have emerged that require real-time, high-performance network communication, such as Cyber-Physical Systems (CPS) and Human-in-the-Loop (HITL) applications. 
These systems exchange information between components, including physical sensors, actuators, and computing devices, to support various applications, from industrial automation to virtual reality. 
To ensure that these applications function correctly and safely, it is necessary to guarantee a certain level of end-to-end network latency. For instance, industrial automation applications require an end-to-end latency of 1 ms with extremely high reliability ($>99.999\%$) \cite{lema2017business}. 
End-to-end latency for typical CPS and HITL applications refers to the time it takes for a packet of data to travel from the sensor source to the controller and from the controller back to the actuator.
In such scenarios, delays in network communication can have serious consequences, such as delays in the response of a robotic system or human feedback.
This requires a deep understanding of the underlying network infrastructure and developing novel techniques to bind latency with extremely high reliability.
For switched Ethernet networks, these requirements can be achieved using mechanisms specified under IEEE 802.1 Time-Sensitive Networking (TSN). 
TSN is a suite of standards that specify mechanisms that enable time-critical traffic alongside best-effort traffic \cite{nasrallah2018ultra}. 
TSN employs various packet scheduling and traffic-shaping mechanisms, e.g., Time-aware Shaping (IEEE 802.1Qbv), to guarantee end-to-end latency. 
However, implementing TSN mechanisms for wireless networks is challenging due to their stochastic nature.


\begin{figure}
\centering
\begin{subfigure}{0.9\linewidth}
    \begin{center}
        \includegraphics[width=1\linewidth]{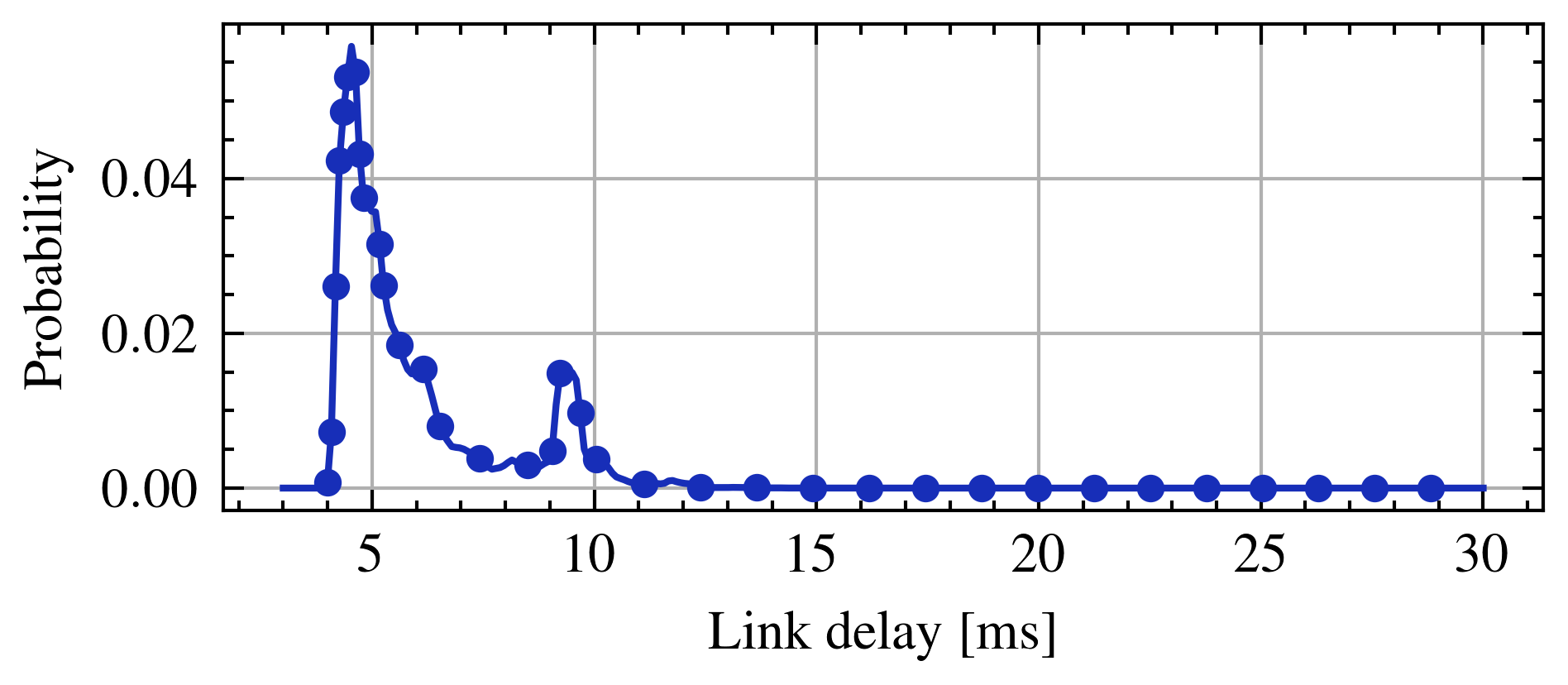}
    \end{center}
\end{subfigure}
\begin{subfigure}{0.9\linewidth}
  \begin{center}
        \includegraphics[width=1\linewidth]{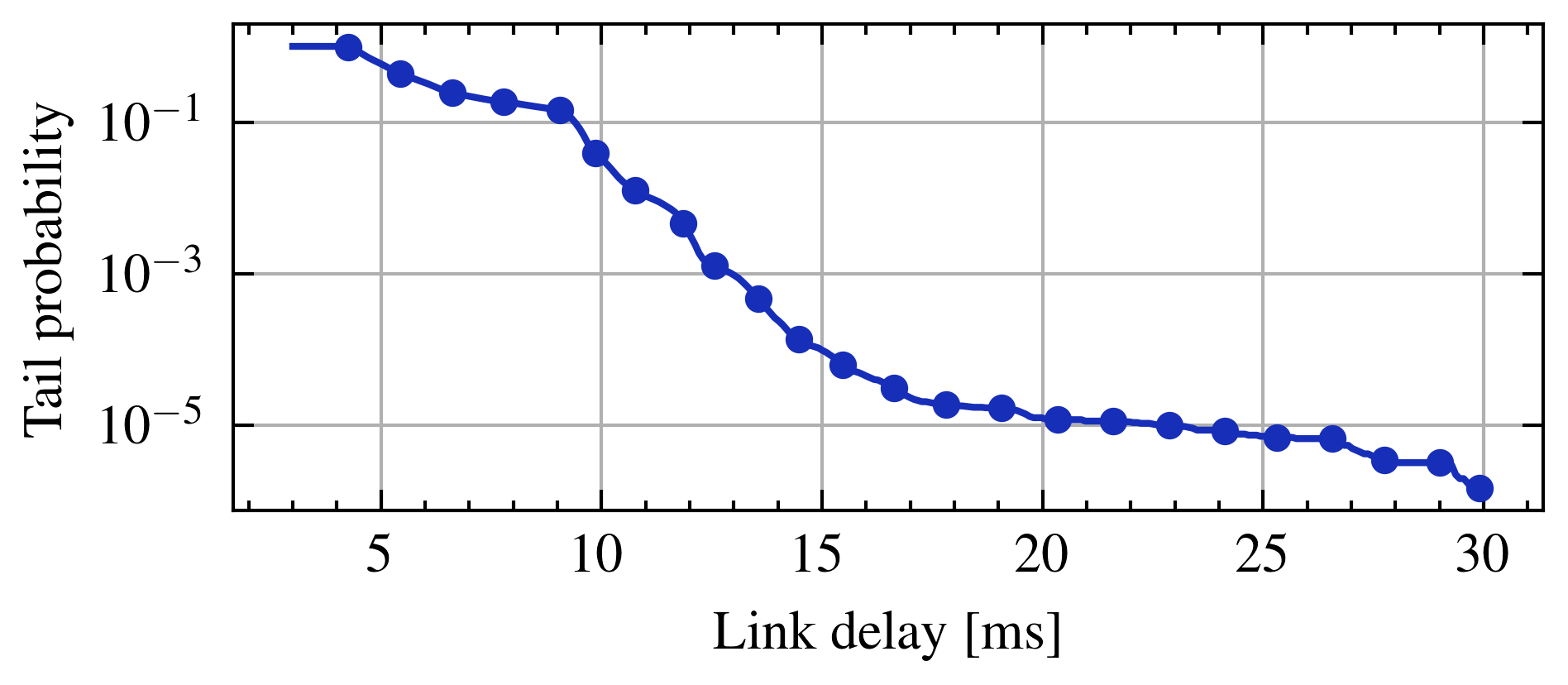}
    \end{center}
\end{subfigure}
\caption{5G downlink measurements depicting the latency probability distribution and tail distribution.}
\label{fig:tail-intro}
\end{figure}

Wireless mobile communication links are susceptible to interference and potentially complicated channel scattering and fading patterns, unlike isolated wired links. 
Hybrid Automatic Repeat Request (HARQ) is used in most wireless communication schemes to ensure reliability by retransmitting lost or corrupted packets. 
It works by combining Forward Error Correction (FEC) with Automatic Repeat Request (ARQ) protocols, allowing for error detection and correction at both the physical and link layers of the communication ~\cite{3gpp.38.321}. 
However, it introduces additional processing and retransmission time for lost or corrupted packets. 
This latency can depend on the specific HARQ implementation and the characteristics of the wireless channel, such as signal strength and interference. 
Ultimately, compared to wired medium, packets traversing the wireless link will experience a non-deterministic latency with 1) a higher average and 2) significant packet delay variation.

Despite the attempts made to reduce average latency in wireless communication technologies, such as Ultra-reliable Low-latency Communications (URLLC) improvements \cite{ansari2022performance}, there is still a considerable gap between the application latency requirements with extreme reliability levels of around 99.9999\% and what state-of-the-art wireless networks offer.
Conforming the wireless link to such a reliability level requires controlling the behavior of extremely rare outliers in the latency probability distribution or the tail of the distribution.
The tail of the latency probability distribution refers to the portion that extends to the far right and contains low-probability latencies, which are less likely to occur than the bulk of the distribution. 
In other words, it represents the extreme values of the distribution as shown in Figure \ref{fig:tail-intro}.
By analyzing the tail of the latency distribution, it is possible to estimate the likelihood of rare latencies, which can be used to make more informed decisions in wireless transmission \cite{sharma2023deterministic}.
All in all, to achieve the requirements, latency must be analyzed using probabilistic models where the accuracy of the tail is of great importance.

\subsection{Related Works}


Latency probability prediction is a crucial component of network performance evaluation; especially, to support time-critical applications over wireless networks. 
In the literature, several approaches to characterize latency in communication networks have been proposed. 
These approaches could be essentially grouped into two categories: (i) model-based approaches and (ii) data-driven approaches. 
In model-based approaches, wireless networks and services are modeled, e.g., as queueing systems, and analytical tools (e.g., network calculus) are applied to derive bounds on latencies. 
For instance, stochastic network calculus was used to obtain probabilistic bounds on the end-to-end delay in a multi-hop wireless network in \cite{champatiTransientAnalysisMultihop2020b}.
In contrast to model-driven approaches, data-driven approaches employ machine-learning methods to identify and learn relationships between latency and other variables using the measurement data from real systems. 
Khangura et al. identified bottleneck links and estimated the residual bandwidth via a neural network that was trained using a vector of packet dispersion values \cite{khanguraMachineLearningMeasurementbased2019}. 
The output of the estimator produced a point estimate of the performance metric (i.e., available bandwidth). 
Point estimates (e.g., average throughput and average delay) are insufficient for scenarios where applications expect very high-quantile performance guarantees (e.g., industrial automation expects a 1 ms delay with a 99.999\% guarantee). 
For such scenarios, complete probability distributions of performance metrics are necessary \cite{bennis2018ultrareliable}.
Within the data-driven approaches proposed in the literature, we focus on those that provide probability distributions for different performance metrics. Using the histogram-based approach, conditional RTT probability distribution in IoT systems is estimated by Flinta et al. in \cite{flintaPredictingRoundTripTime2020}. 
In \cite{samaniConditionalDensityEstimation2021}, Samani et al. estimated the conditional probability distributions of network performance metrics (i.e., frame rate and response time) based on the measurements taken from a test network. 
It was observed that \glspl{MDN} (Gaussian and Log-normal) and histograms could be leveraged to generate conditional distributions with their respective trade-offs. 
Similarly, authors in \cite{sawabeDelayJitterModeling2022} proposed to use a mixture of Laplace distributions to characterize delay jitter in high-frequency and mobile communications. 
It is worth pointing out that these estimators are apt for predicting the bulk of latency distribution but not tail probabilities, which could not be ignored for applications demanding strict latency guarantees. 
In our previous work, the end-to-end latency in a multi-hop queuing system was characterized by exploiting Extreme-value Theory (EVT) in combination with \glspl{MDN} \cite{samie_sec_2021}.
The evaluation results showed the superiority of this method in predicting tail probabilities, over the traditional \gls{MDN} approaches for conditional latency probability prediction. In this paper, we compare the performance of these estimators in predicting the latency probabilities of wireless networks.  

\subsection{Contributions}
The main contribution of our work is a data-driven approach to characterize the latency of wireless networks in a probabilistic way with a focus on the extreme latencies, i.e., tail probabilities.
Unlike model-driven approaches, we do not assume the wireless channel model (e.g., Rayleigh fading model) or any traffic profile (e.g., Poisson packet arrival).
We investigate the performance of the state-of-the-art probability density estimation approaches, namely \glspl{MDN}, specifically on the extremely rare latencies that occur with probabilities as small as $10^{-6}$.
In our previous work, we show how EVT models can be incorporated into \glspl{MDN} to accurately relate the tail latency distributions to the conditions of the network. 
This approach is compared to the state-of-the-art methods through an extensive evaluation of two actual implementations of the 5G network and an implementation of IEEE 802.11 network: (i) \gls{COTS} 5G, (ii) OpenAirInterface (OAI) 5G, and (iii) Mangocomm IEEE 802.11 (WiFi)\footnote{https://www.mangocomm.com/802.11-mac-phy}.
The evaluations reveal the proposed approaches' advantages, disadvantages, and sensitivities when characterizing the tail latency behavior (e.g., at $> 99.999\%$ quantiles).

\section{System Model and Problem Statement}

As mentioned before, we strive to build a probabilistic understanding of the delay of the packets sent over the 5G network which is a bridge in the \gls{TSN} domain, or interfaces the end node with a controller.
Assume packets $\{1,...,N\}$ are being pushed to the wireless network with a fixed length $B_s$ and at a fixed interval $T_s$ that mimics the IEEE 802.1Qbv time cycles or sensor/actuator period packets arrival profile \cite{craciunas2016scheduling}.
We denote the observed end-to-end delay of the packet $n$ by $Y_n$ and the transmission conditions by $X_n$.
Due to the nature of the wireless medium, packets may need to be re-transmitted to ensure reliability.
Such mechanism is implemented in \gls{HARQ} scheme which is widely used in the majority of wireless networks.
These random re-transmissions which are the main contributors to the latency, could be related to the transmission conditions such as \gls{SINR} and \gls{MCS} index that we model in $X_n$.
Let the random variable $Y$  with domain $\mathcal{Y} \subset \mathbb{R}_{+}$ model the packets' latency, its probability density could be expressed conditioned on the transmission conditions i.e. $\mathbb{P}\left[Y \mid X = \boldsymbol{x}\right]$ where $\boldsymbol{x}$ denotes the observed network state or condition.

\textbf{Problem Statement: }
We are interested in predicting the probability density of latency from the network conditions, i.e., the probability that the packets face a certain delay, given the transmission conditions of the network, i.e., 
\begin{equation} \label{eq:prob}
  \hat{p}\left(Y \mid X = \boldsymbol{x}\right) \approx \mathbb{P}\left[Y \mid X = \boldsymbol{x}\right].
\end{equation}
In particular, the accuracy of $\hat{p}$ is of great importance for large values of $Y$ or at the tail of the probability distribution.
The tail probability is then characterized by $1-\hat{p}\left(Y > y \mid X = \boldsymbol{x}\right))$.

\section{Approach}

The above task of estimating the wireless link's latency probabilities from the transmission conditions falls into the domain of \gls*{CDE}.
In a non-conditional density estimation scenario, we can fit a parametric model such as a \gls*{GMM} to the set of latency samples.
This parametric density function is described by a finite-dimensional parameter $\theta$~\cite{Bishop94mixturedensity}. 
For example, $\theta$ could be a vector of the weights, locations, and variances of the \gls*{GMM} density function.
In \gls*{CDE}, we need to devise a function denoted by $h_{\omega}$ to map the values of $X$, i.e., conditions, into the space of the density function parameters $\theta$ in the form $\theta = h_{\omega}\left(\boldsymbol{x}\right)$.
\Gls*{MDN} is a well-known method that uses a fully connected neural network as $h_{\omega}$ to control the parameters of the conditional density estimate $\hat{p}_{\theta}$ \cite{Bishop94mixturedensity}.
In \gls*{MDN}, \gls*{MLE} is used for estimating the parameters $\omega$, i.e., $\omega$ is chosen so that the conditional likelihood of the samples $\mathcal{D} = \{ (\boldsymbol{x}_{1},y_{1}),...,(\boldsymbol{x}_{N},y_{N}) \}$ is maximized.
This is equivalent to minimizing the \gls*{KL-divergence} between the empirical data and the parametric density $\hat{p}_{\theta}$ \cite{Bishop94mixturedensity}.

\begin{figure}[t!]
  \centering
    \includegraphics[width=0.99\linewidth]{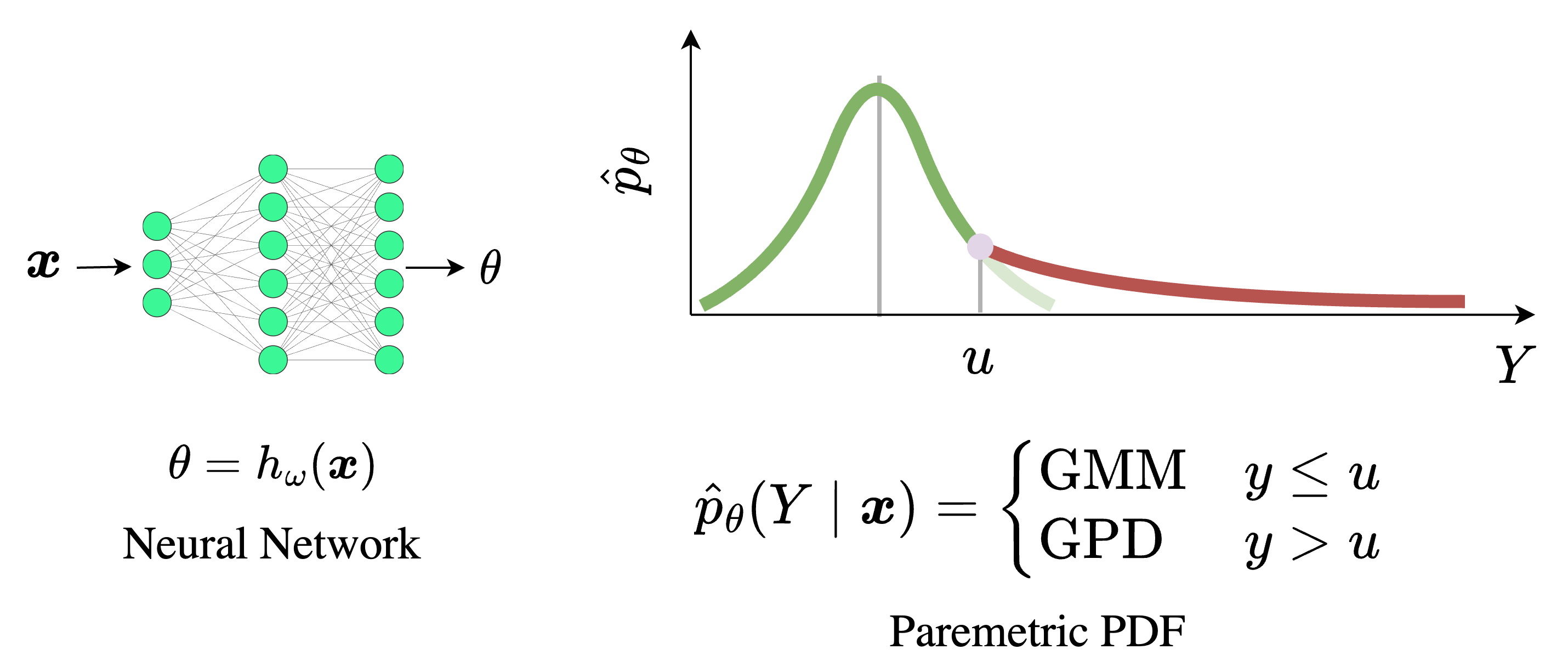}
  \caption{Block diagram of the predictor, showing inputs, components, and outputs.}
  \label{fig:approach}
\end{figure}

The subsequent procedure involves determining the parametric distribution $\hat{p}_{\theta}$ employed in the \gls*{MDN}. 
In this regard, we primarily use \gls*{GMM}, a widely recognized and established parametric distribution, in addressing the requirements of this particular problem domain.
However, our use case requires high accuracy at the tail of the fitted probability density.
In our previous work \cite{samie_sec_2021}, we investigated the use of \Gls*{EVT} models for the tail estimation in addition to the Gaussians for this task.
Since the output of the \gls*{GMM}-based solutions is a density function whose tail probability decreases exponentially fast, it can lead to significant probability error for low probability events,  e.g., in the order of $10^{-5}$ if the empirical distribution is not light-tailed.
we use the following parametric mixture function that encapsulates the \gls*{GPD} tail model in combination with the \gls*{GMM} to be fused with the neural network of a \gls*{MDN} in the form:
\begin{equation}
    \hat{p}(y|\theta = h_{\omega}\left(\boldsymbol{x}\right)) =
    \begin{cases}
        f(y|\phi) & y \leq u \\
        [1-F(u|\phi)]g(y|\beta,\xi,u) & y > u,
    \end{cases} \label{eq:mix}
\end{equation}
where $f(y|\phi)$ and $F(y|\phi)$ denote the \gls*{GMM}'s \gls*{PDF} and \gls*{CDF} respectively, $g(y|\beta,\xi,u)$ denotes the \gls*{GPD}'s \gls*{PDF}, $u$ is the tail threshold, and $\theta$ is the collection of parameters $\phi$, $\beta$, $\xi$, and $u$.
The structure of the predictor is illustrated in Figure \ref{fig:approach}.
In this extreme value \gls*{MDN} scheme, tail parameters of the distribution (threshold, tail index, and scale) are being estimated by the neural network in addition to the bulk distribution parameters.

\section{Methodology}

To evaluate the proposed approach, we attempt to predict the latency of a real 5G network, i.e., $Y$, under different network conditions, i.e., $X$.
For this task, first, we collect latency samples from the network to form a dataset that can be used to train and evaluate the \gls{MDN} model.
Then, we compare these models in terms of their accuracy in predicting the tail latency distribution.
Other aspects such as the sensitivity to the number of training samples, noise regularization, and generalization accuracy are investigated as well.
The following subsection introduces all the steps for benchmarking the predictors and conducting the experiments. \footnote{The reproducible results and the datasets could be found at: https://github.com/samiemostafavi/wireless-pr3d}.

\begin{figure}[h]
    \includegraphics[width=0.99\linewidth]{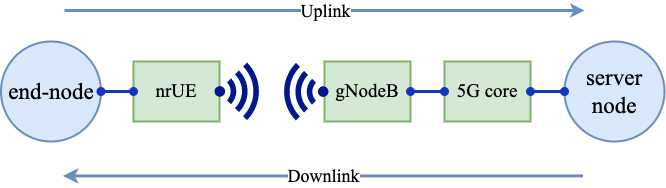}
    \caption{Illustration for the setup used to gather measurement data from both COTS 5G and OAI 5G.}
    \label{fig:measurement-setup}
\end{figure}

\begin{figure}[h]
    \includegraphics[width=0.99\linewidth]{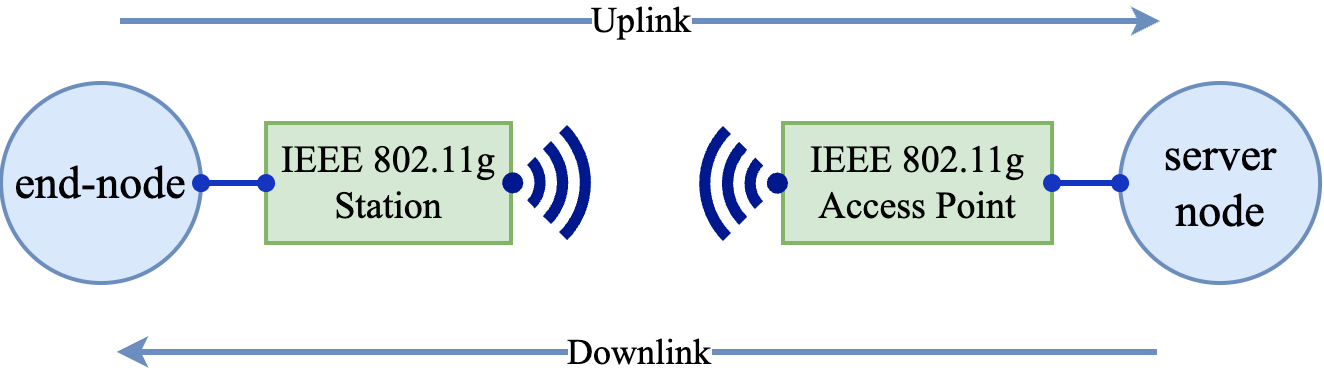}
    \caption{Illustration for the setup used to gather measurement data from WiFi network.}
    \label{fig:wifi-measurement-setup}
\end{figure}

\textbf{Measurements Setup}: 
The experimental setup consists of 2 nodes connected using a 5G network as shown in Figure \ref{fig:measurement-setup} or WiFi network as shown in Figure \ref{fig:wifi-measurement-setup}.
By default, the latency samples are collected from 10ms, 172 bytes periodic transmission of timestamped packets from the end-node to the server node (uplink) and back (downlink).
Since the nodes reside on different machines, we synchronized their clocks using PTP protocol on a dedicated separate interface.

The desired network condition is also recorded for each latency sample.
We conduct the 5G measurements on two different systems: a private \gls{COTS} 5G, and an end-to-end \gls{SDR} 5G developed by OAI software alliance ~\cite{oai5g_2019}.
The \gls{SDR} 5G system offers maximum flexibility in recording and manipulating the parameters of all layers. 
Both 5G networks operate in band 78, in \gls{TDD} mode, with 106 \glspl{PRB} which occupies 40 MHz of bandwidth at 3.5 GHz.
The WiFi network operates on 2.4GHz frequency with 20MHz of bandwidth on the same location and \glspl{SDR} as \gls{SDR} 5G.

In the \gls{COTS} 5G experiment, we gathered 4 million uplink latency measurements in 11 hours, from 20 distinct locations within a 50 $\text{m}^{2}$ 5G-covered conference room containing metallic chairs, whiteboards, and screens.
The 5G base station was located in one corner of this rectangular-shaped room.
The investigation did not reveal any variations in the tail behavior across diverse locations.
Therefore, we analyze non-conditional latency prediction on the \gls{COTS} 5G measurements.
This approach is viable in cases where there is no information about the communication conditions, or there is no relation between the available conditions and the latency distribution.
Examining the impact of sample size and the parametric model on the fitting accuracy remains essential, irrespective of the neural network's inherent bias-variance trade-off.
In the \gls{SDR} 5G experiment, we gathered 5 million uplink latency measurements in 15 hours.
In the configuration employed, we located the base station and \gls{UE} 1 meter apart and in line of sight.
We were able to set the \gls{MCS} index of the transmissions to a fixed value of 3, 5, or 7 throughout an entire round, in contrast to the link adaptation algorithm, which assigns a value of 15 in the absence of such a constraint.
In this case we analyze conditional latency prediction approach.

Finally, in the WiFi experiment, we gathered 4 million uplink latency traces from round-trip measurements in 3 hours.
The access point and station were located 1 meter apart and in line of sight as in \gls{SDR} 5G case.
We set the packet size for an entire round to different values of 172 bytes, 3440 bytes, 6880 bytes, and 10320 bytes.
In this case, again we analyze conditional latency prediction approach.


\textbf{Data Preprocessing}: 
We observed that normalization and standardization of the data play a crucial part in achieving the best accuracy in training.
For instance, the delay values are scaled to the millisecond scale and moved to around zero by subtracting the average delay.
Conditions are normalized to be between 0 and 1.
Moreover, applying noise regularization in mixture density networks is suggested by \cite{rothfuss2020noise} to improve the accuracy of the fit.
We investigate that by comparing trained models on two noise levels of 1ms and 3ms with the raw data.  




\textbf{Training}: 
The \gls{MDN} model is implemented using four fully connected hidden layers with [10, 100, 100, 80] neurons, respectively. It controls parameters of 15 Gaussian centers and optionally a \gls{GPD} for modeling the tail.
The output size of the neural network is equal to the number of parameters of the parametric density function $\hat{p}_{\theta}$, which is 48 when \gls{GPD} is incorporated and 45 without \gls{GPD}.
We name the model with \gls{GPD} component, \gls{GMEVM}.
Training the model is done by Adam optimizer in 4 rounds, each with 200 epochs and different learning rates: $[10^{-2}, 10^{-3}, 10^{-4}, 10^{-5}]$, respectively.
The batch size is set to 1/8 of the training dataset size.
We observed that including training rounds with minimal learning rates, e.g., $10^{-4}$ or $10^{-5}$ is necessary to avoid underfitting in the tail region.
In Table \ref{tab:training-dur}, the training durations for different sizes of the training dataset are mentioned.
\Gls{MDN} models are implemented and evaluated using CPU-only Tensorflow on a system equipped with Intel(R) Core(TM) i9-10980XE CPU operating at 3.00GHz.
\begin{table}[h!]
  \vspace{0.05in}
  \begin{center}
    \caption{Training durations for 1000 epochs}
    \label{tab:training-dur}
    \begin{tabular}{|m{2.5cm}|m{1cm}|m{1cm}|m{1cm}|}
    \hline
    Training dataset size & 1M & 256k & 64k \\
    \hline
    Batch size  & 128k & 32k & 8k \\
    \hline
    Epoch duration  & 4s & 1s & 300ms \\
    \hline
    Step duration & 500ms & 100ms & 27ms \\
    \hline
    Total training time & 66.6m & 16.6m & 5m \\
    \hline
    \end{tabular}
  \end{center}
\end{table}

The \gls{MDN} model estimates the PDF of the latency distribution from the observed latency measurements over time.
In a slowly changing wireless environment, we can periodically collect new latency measurements and use them to re-train the \gls{MDN} model.
Therefore, it is crucial to carefully monitor the model's performance concerning the retraining frequency, which is limited by the training duration and dataset size.


\textbf{Evaluation}: 
In our study, we assess the efficacy of our trained models by conducting training sessions for ten different models. 
We then use the obtained results to compute the minimum, maximum, and average predictions. 
Subsequently, we compare these predictions to the ground truth, which we established by measuring the system's latency.
We collect many samples to ensure the accuracy of our ground truth measurements. 
To visualize this, we use a colored area between the minimum and maximum curves instead of only one.
We consider a predictor to be effective when its average prediction aligns closely with the ground truth and exhibits lower variance.
This approach enables us to gain a deeper understanding of the performance of our models and the extent of the variations.

\section{Numerical Results}

This section focuses on analyzing the performance of latency predictors in various practical scenarios and wireless communication technologies. 
The primary objective is to evaluate the sensitivity of these predictors to the number of samples and the complexity of the data. 


\subsection{Non-Conditional \gls{COTS} 5G Latency Prediction}
\label{sec:non-cond-pred}

\begin{figure}[ht!]
\centering
\includegraphics[width=0.8\linewidth]{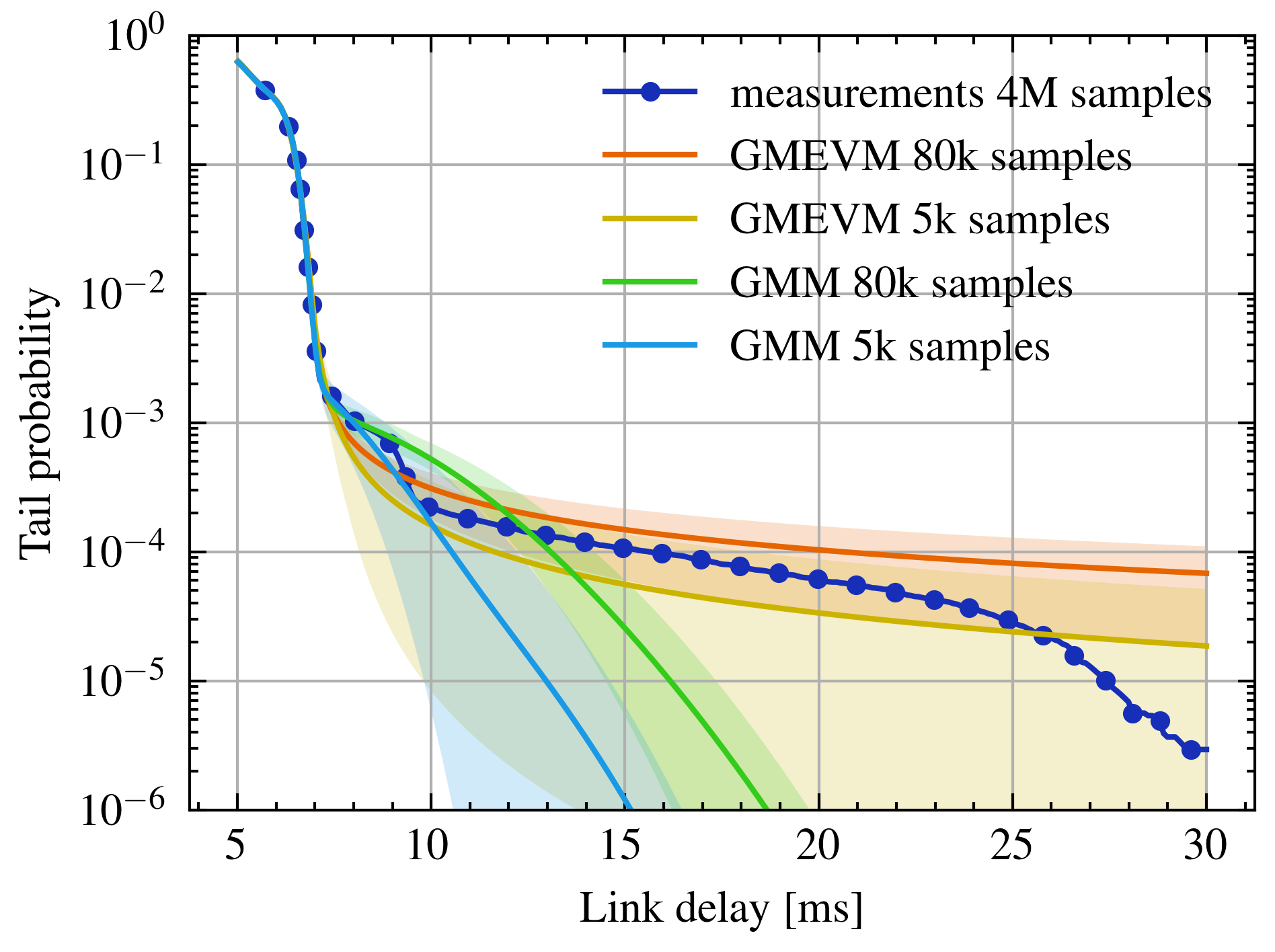}
\caption{\gls{COTS} 5G uplink latency measurements vs parametric density fits with different number of samples and models}
\label{fig:ep5g-tail}
\end{figure}

\begin{figure}[h!]
\centering
\includegraphics[width=0.8\linewidth]{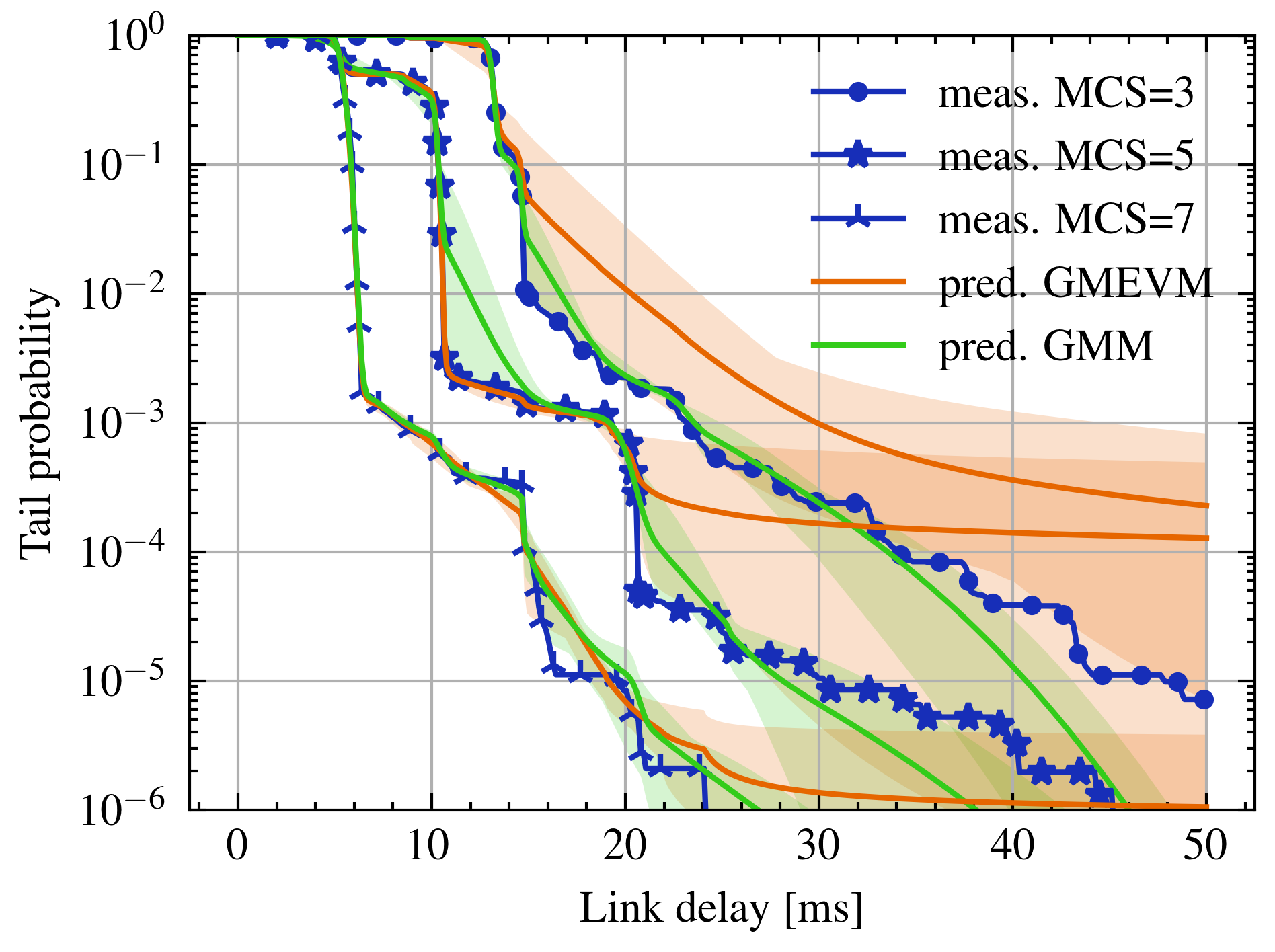}
\caption{Performance of \gls{MDN} models trained without noise regularization. Number of samples: 1M (20\%)}
\label{fig:oai5g-nonoise}
\end{figure}

\begin{figure*}[ht!]
  \vspace{0.05in}
  \begin{subfigure}{0.33\linewidth}
    \centering
    \includegraphics[width=1\linewidth]{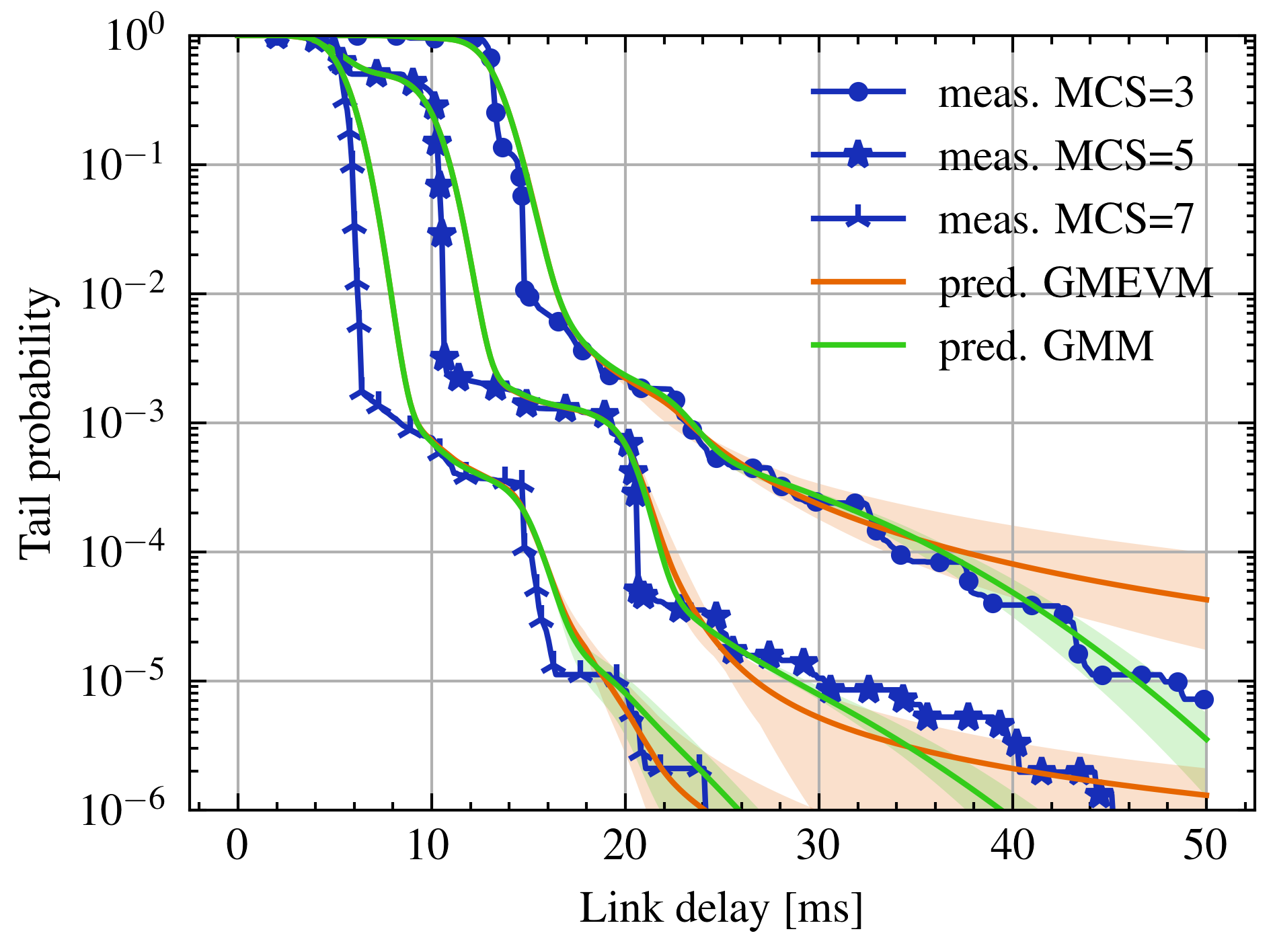}
    \caption{Models trained with 1M samples (20\%)}
    \label{fig:oai5g-tail:a}
  \end{subfigure}
  \begin{subfigure}{0.33\linewidth}
    \centering
    \includegraphics[width=1\linewidth]{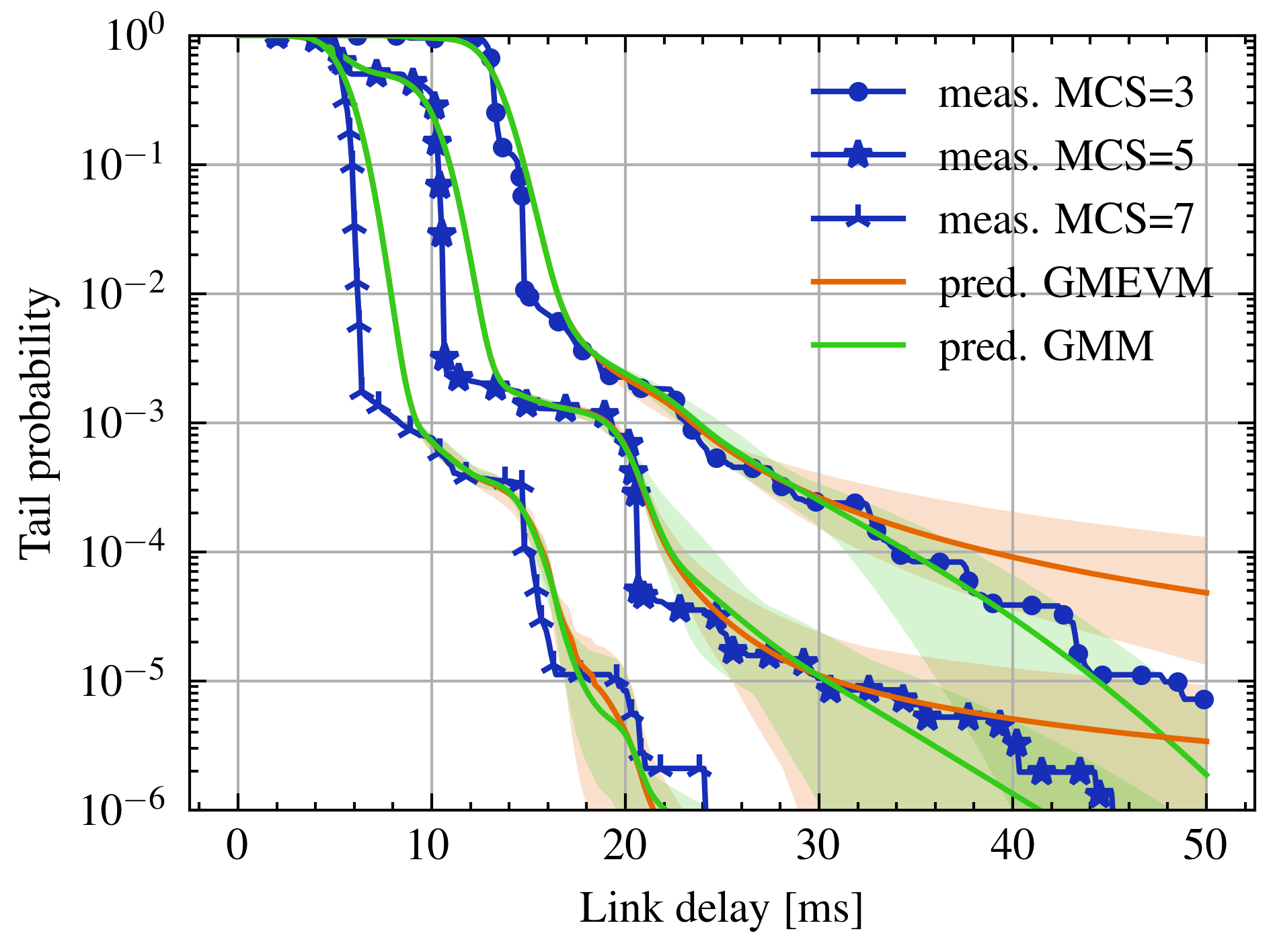}
    \caption{Models trained with 250k samples (5\%)}
    \label{fig:oai5g-tail:b}
  \end{subfigure}
  \begin{subfigure}{0.33\linewidth}
    \centering
    \includegraphics[width=1\linewidth]{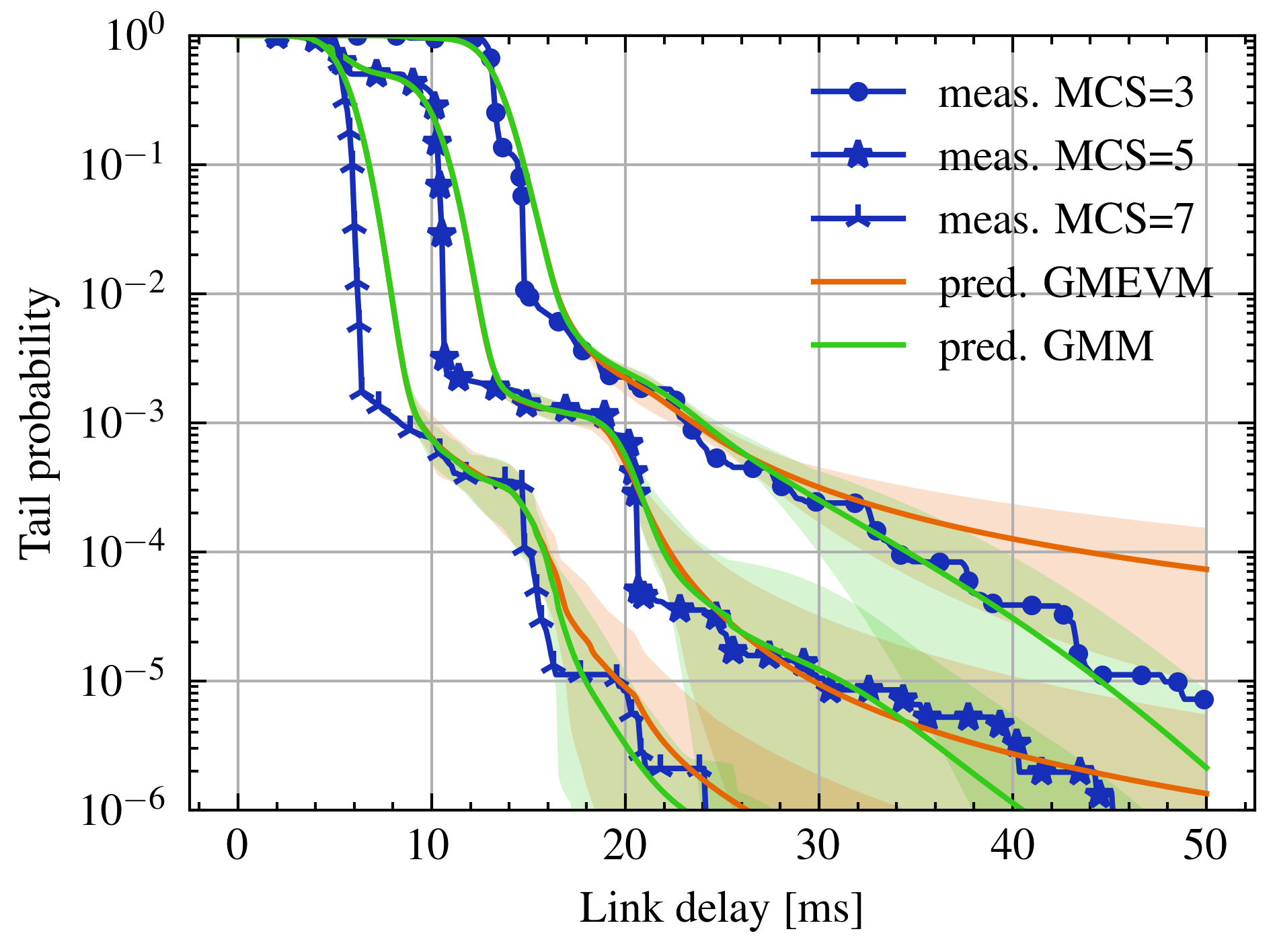}
    \caption{Models trained with 64k samples (1.3\%)}
    \label{fig:oai5g-tail:c}
  \end{subfigure}
  \caption{\gls{SDR} 5G uplink latency tail probability measurements conditioned for different \gls{MCS} indices with 5M samples vs predictions of \gls{MDN} models trained with different numbers of samples and added Gaussian noise}
  \label{fig:oai5g-tail}
\end{figure*}

In a relatively simple approach, prediction of the latency probability could be carried out only from the samples and not incorporating any condition.
Figure \ref{fig:ep5g-tail} shows the tail probability estimation of the \gls{COTS} 5G latency for different models and number of samples. 
It can be observed that the tail latency distribution of the \gls{GMM} predictor falls exponentially with the received delay and therefore diverges from the empirical distribution whereas, the \gls{GMEVM} predictor closely follows the empirical distribution.
Next, we compared the impact of training data size on the accuracy of the fit for \gls{GMM} and \gls{GMEVM}. 
It can be expected for both predictors that their performance improves with the increasing number of training data samples. 
Our observations indicate that the effect of decreasing the sample size from the magnitude of $10^{5}$ to $10^{4}$ on the \gls{GMEVM} is relatively minor. 
However, the accuracy is significantly compromised when reducing the sample size to $10^{3}$, as depicted in the figure.
The emergence of a distinct trend in the measurement curve as the tail probability approaches the order of $10^{-4}$ could explain the observed behavior. 
Our findings reveal that \gls{GMEVM} can effectively capture this trend, whereas \gls{GMM} cannot.
In the next section, we assess the performance of the \glspl{MDN} in the context of conditional scenarios and a more challenging latency profile.

\subsection{Conditional \gls{SDR} 5G Latency Prediction}

This section investigates the performance of \glspl{MDN} in the context of conditional latency prediction. 
Condition vector $X$ is related to the parametric distribution parameters $\theta$ through the \gls{MLP}.
It can potentially extrapolate the observed behavior of the training conditions to previously unseen ones.
As a result, it may be possible to leave certain conditions unsampled and rely on the MLP's ability to predict their profile, further reducing the number of training samples while achieving a desirable level of accuracy. 

In this context, it is expected that the link capacity will increase with the increase in the \gls{MCS} index when the index is less than 15.
Furthermore, an increase in the link capacity resulting from using a higher \gls{MCS} index is anticipated to decrease the delay.
Figure \ref{fig:oai5g-nonoise} depicts the measured tail probabilities for different \gls{MCS} indices using all 5 million samples (1.6 million samples for each condition).
Moreover, we trained \gls{GMM} and \gls{GMEVM} models using 1 million samples, including all three \gls{MCS} conditions.
In the presented results, the \gls{GMEVM} scheme exhibits high variance and poor prediction accuracy across all three cases, while \gls{GMM} performs well, as evidenced in the figure. 
We hypothesize that the measurements' bumpy and non-smooth tail profile is the reason for the observed struggles of \gls{GMEVM}. 
To address this, we propose a noise regularization approach by adding random Gaussian noise with a variance of 1 millisecond to the latency samples. 
As shown in Figure \ref{fig:oai5g-tail}, this technique significantly improves the performance of \gls{GMEVM} even with lower training samples by smoothing out the tail.
However, introducing noise regularization comes with a cost, which is lower accuracy in the bulk of the latency distribution.
It is crucial to select the appropriate variance for the added noise.
\begin{figure}[ht!]
\centering
\includegraphics[width=0.8\linewidth]{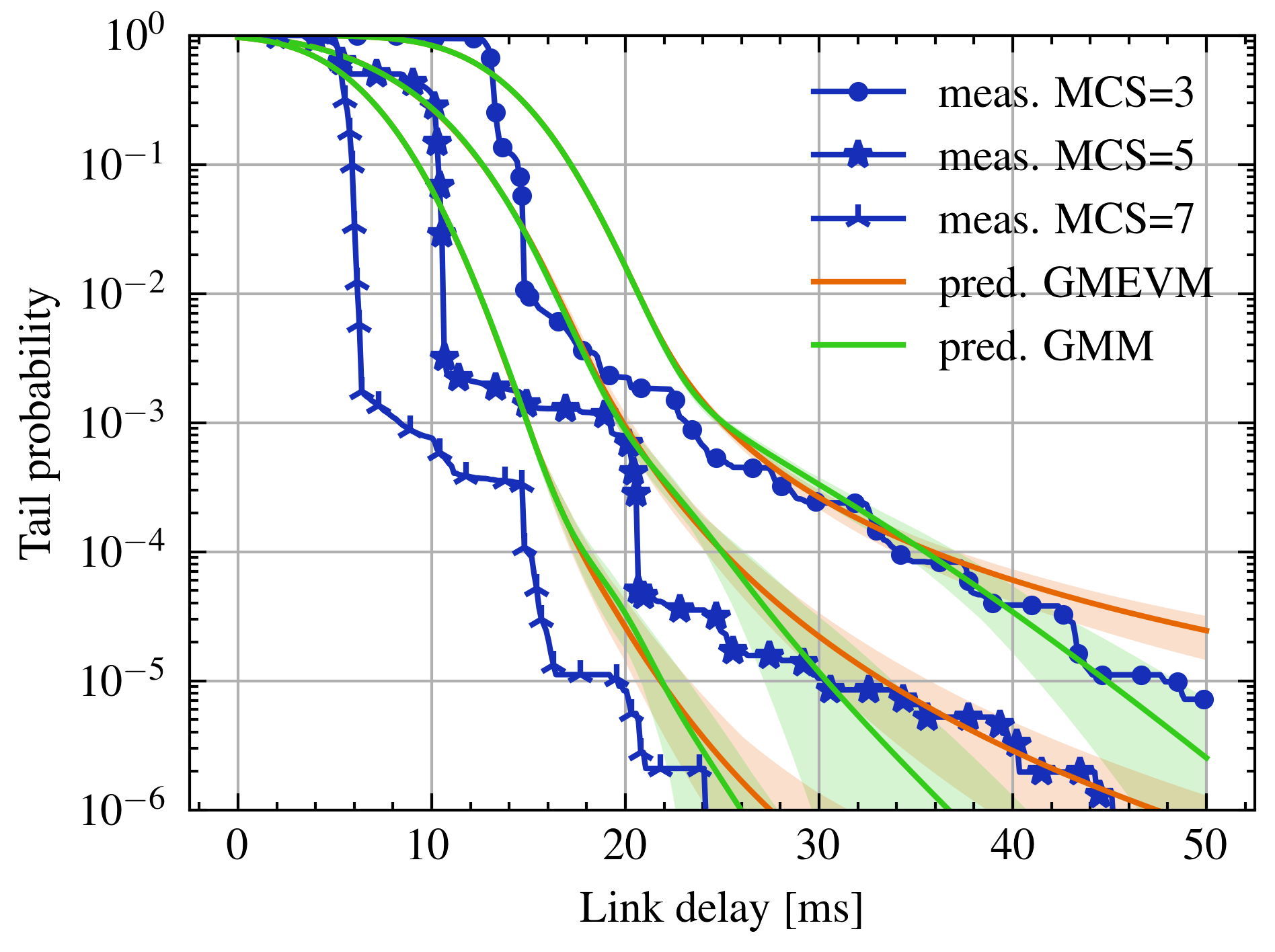}
\caption{Performance of \gls{MDN} models trained with added Gaussian noise of variance 3ms. Number of samples: 250k samples (5\%)}
\label{fig:oai5g-verynoisy}
\end{figure}
Figure \ref{fig:oai5g-verynoisy} demonstrates the impact of using a larger variance of 3 milliseconds. 
The noise has degraded the low-tail regions' prediction accuracy but improved the high-tail predictions. 
Therefore, choosing the correct variance value is essential for noise regularization in \gls{GMEVM}.
Figure \ref{fig:oai5g-tail} also shows the models' accuracy concerning the training dataset's size.
We observe higher uncertainty and error from the models in the case of less number of training samples.

\begin{figure}[ht!]
\centering
\includegraphics[width=0.8\linewidth]{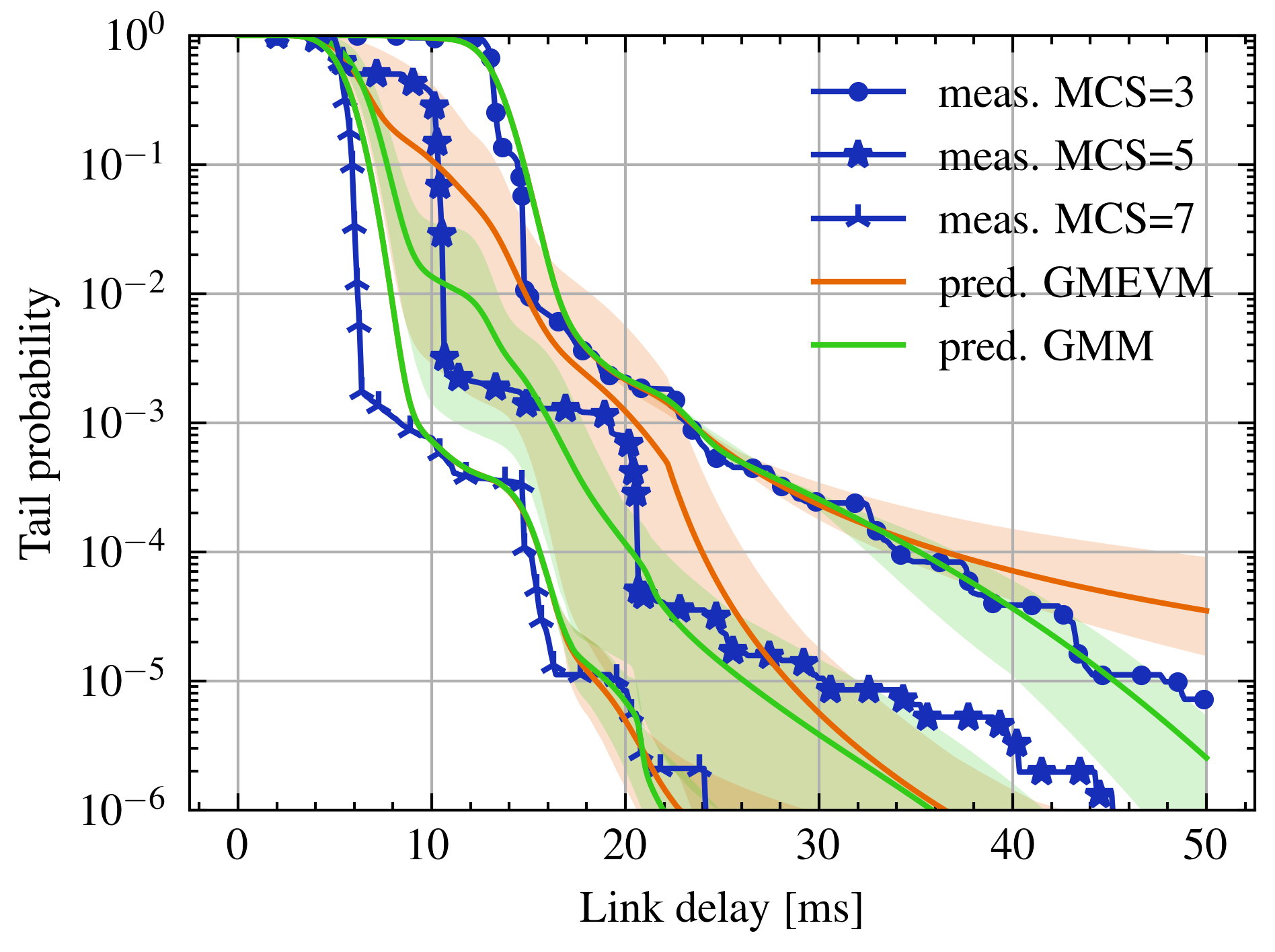}
\caption{Generalization capability of \gls{MDN} models trained without MCS=5 samples. Number of samples: 0.6M (20\%)}
\label{fig:oai5g-no5}
\end{figure}
In Figure \ref{fig:oai5g-no5}, we evaluate the generalization capability of \gls{MDN} models. 
For this purpose, we trained the models on a dataset that does not include latency samples with \gls{MCS}=5.
Consequently, \gls{MCS}=5 is an unseen condition for the predictors. 
Both models follow the empirical tail but with a relatively high variance. 
\gls{GMM} exhibits better performance than \gls{GMEVM} with a lower variance.
One can see an opportunity for further improving the performance of the models in handling unseen conditions in future work.

\subsection{Conditional WiFi Latency Prediction}

\begin{figure}
\centering
\begin{subfigure}{1\linewidth}
    \begin{center}
        \includegraphics[width=0.85\linewidth]{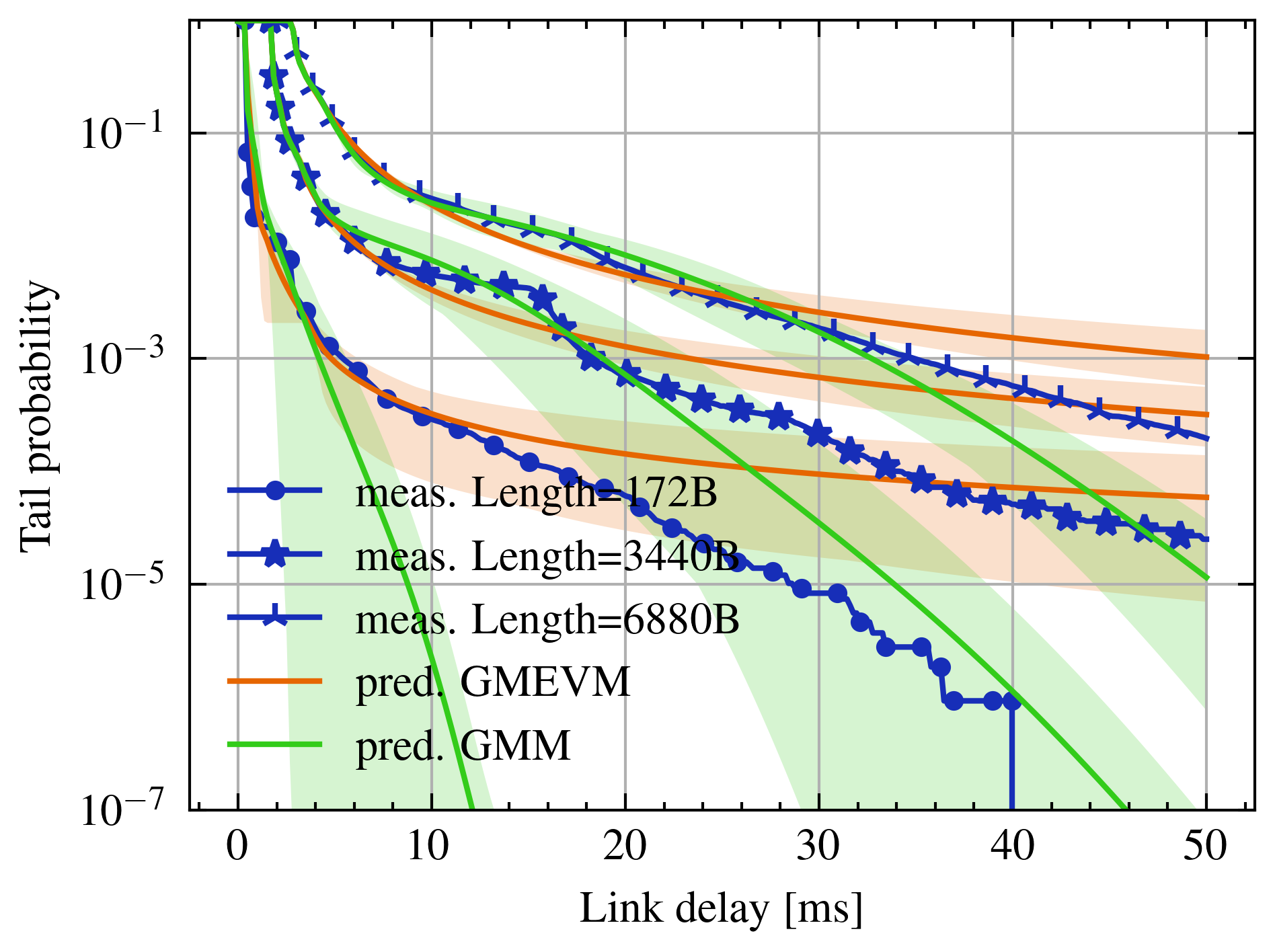}
    \end{center}
    \caption{Models trained with 32k samples (3.2\%)}
    \label{fig:wifi-tail-l:a}
\end{subfigure}
\begin{subfigure}{1\linewidth}
  \begin{center}
        \includegraphics[width=0.85\linewidth]{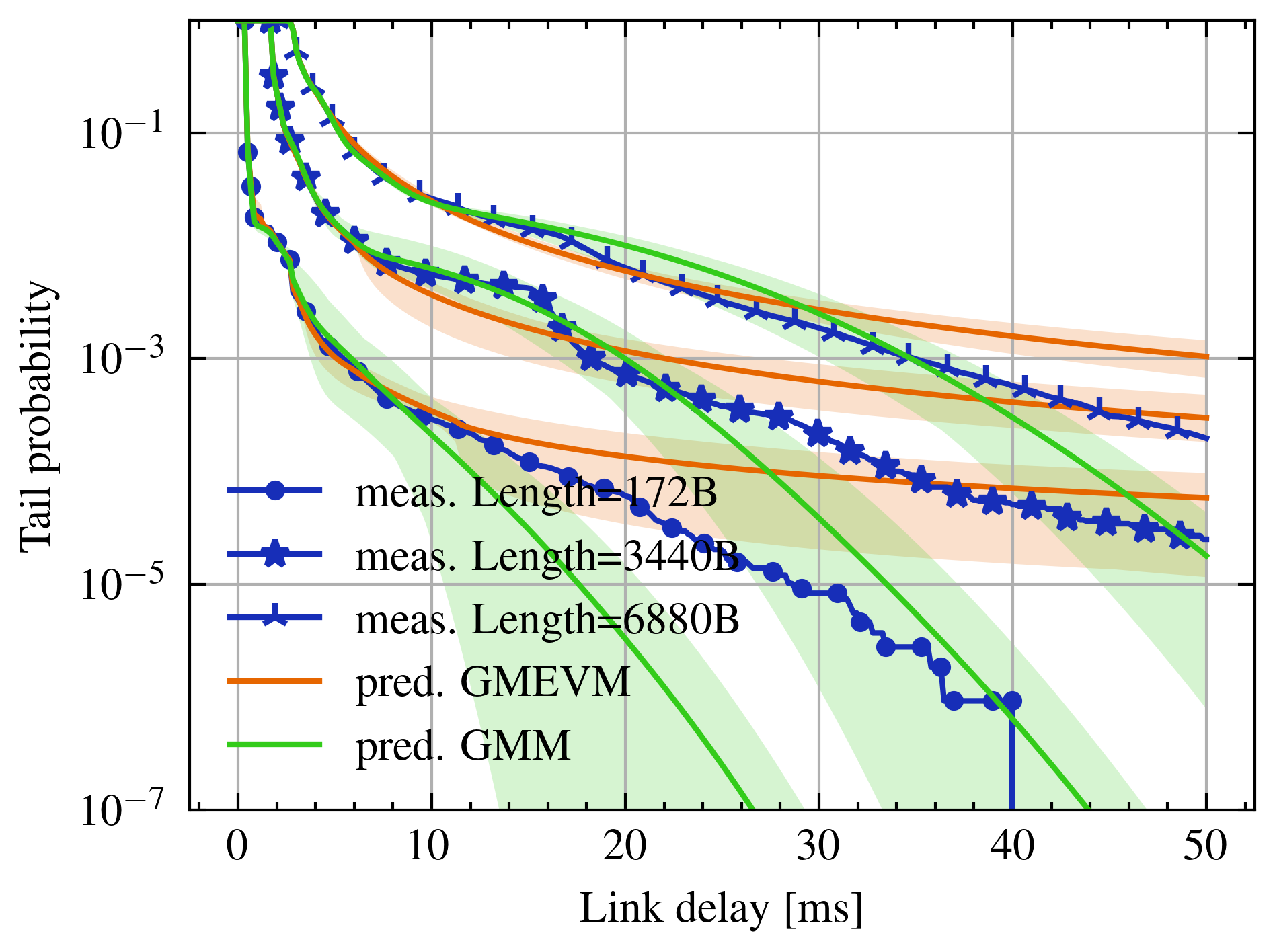}
    \end{center}
    \caption{Models trained with 512k samples (50\%)}
    \label{fig:wifi-tail-l:b}
\end{subfigure}
\caption{\gls{SDR} WiFi downlink latency tail probability measurements conditioned for different packet lengths vs predictions of MDN models trained with different numbers of samples.}
\label{fig:wifi-tail}
\end{figure}

In the last experiment, we used the load intensity as a condition on the latency probabilities. 
We gathered 1M latency measurement samples from the SDR WiFi setup for three different packet length values: 172B, 3440B and 6880B. 
As shown in Figure \ref{fig:wifi-tail}, as the packet length increases, the tail of the empirical distribution becomes heavier as a result of higher PDV from the increased queueing and service delays. 
The tail probability predictions obtained from both \gls{GMM} and \gls{GMEVM} exhibit improved performance with increased training samples. 
However, these predictions diverge significantly from the empirical distribution for higher latency values, despite using 512k training samples. We observed that noise regularization does not improve the prediction, as the tail profile is already smooth enough. 
Hence, we leave the task of enhancing the prediction accuracy to future research endeavors.

\section{Conclusions}
This study uses state-of-the-art data-driven approaches, such as mixture density networks, to predict the latency of wireless links, particularly for extreme latencies that impact time-critical applications. 
Specifically, we analyze Gaussian mixture models and a novel approach that integrates extreme value models into the mixture of parametric distributions. 
Through our investigation, we examine the impact of the number of training samples, the complexity of the tail profile, and the generalization capabilities of these approaches. 
Our results demonstrate that both approaches achieve acceptable accuracy with sufficient training samples.
Additionally, we find that noise regularization improves the accuracy of the fit, particularly in the case of GMEVM when the tail profile is non-smooth. 
There is room for improvement in the generalization capabilities which can be addressed in future research.
Overall, our study sheds light on the effectiveness of data-driven approaches in predicting wireless link latency and highlights areas for further improvement.

\bibliographystyle{ieeetr}
\bibliography{main.bib}

\end{document}